\documentclass[preprint2]{aastex}
\usepackage{natbib}
\usepackage{graphicx}
\usepackage{color}
\def\apj{ApJ}
\def\aj{AJ}
\def\araa{ARA\&A}
\def\aap{A\&A}
\def\mnras{MNRAS}
\def\prd{Phys. Rev. D}
\def\apjs{ApJS}

\shorttitle{A Massive elliptical in extended gravity}
\shortauthors{ M. A. Jim\'enez, G. Garcia and X. Hernandez}

\begin{document}

\title {The massive  elliptical galaxy NGC 4649 from the perspective of extended gravity} 
\author { M. A. Jim\'enez\altaffilmark{1}, G. Garcia, X. Hernandez}
\affil{Instituto de Astronom\'{\i}a, Universidad Nacional Aut\'{o}noma de M\'{e}xico,
Apartado Postal 70--264 C.P. 04510 M\'exico D.F. M\'exico}

\altaffiltext{1}{email:mjimenez@astro.unam.mx}

\begin{abstract}
Elliptical galaxies are systems where dark matter is usually less necessary to explain observed dynamics than in the case of spiral galaxies,
however there are some instances where Newtonian gravity and the observable mass are insufficient to explain their observed structure 
and kinematics. Such is the case of NGC 4649, 
a massive elliptical galaxy in the Virgo cluster for which recent studies report a high fraction of dark matter, $0.78$ at $4R_e$. However this galaxy
has been studied within the MOND hypothesis, where a good agreement with the observed values of velocity dispersion is found. Using a MONDian
gravity force law, here we model this galaxy as a self-consistent gravitational equilibrium dynamical system. This force law reproduces the MOND
phenomenology in the $a<a_{0}$ regime, and reduces to the Newtonian case when $a>a_{0}$. Within the MONDian  $a<a_{0}$ scales,
centrifugal equilibrium or dispersion velocities become independent of radius, and show a direct proportionality to the fourth root of the total baryonic mass, 
$V^{4}\propto(M G a_{0})$. We find that the recent detailed observations of the surface brightness profile and the velocity dispersion profile for this 
galaxy are consistent with the phenomenology expected in MONDian theories of modified gravity, without the need of invoking the presence of any hypothetical 
dark matter.
\end{abstract}


\section{Introduction} 
The flat rotation curves of spiral galaxies have been well known for decades, e.g., \citet{Bosma1981, Rubin1982},
assuming the validity of the Newtonian law of gravity throughout, this results in a discrepancy between the dynamical mass and the luminous mass 
of spiral galaxies. Other types of galaxies exhibit mass discrepancies as well, the most remarkable is the case of dwarf spheroidal galaxies
for which the velocity dispersion is well known and the mass inferred from their internal dynamics greatly exceeds the visible mass by factors 
reaching into the thousands e.g., \citet{Simon2007}. 
On the other hand, bright giant elliptical galaxies exhibit small mass discrepancies e.g., \citet{Romanowsky2003}, usually
these facts are commonly interpreted as the manifestation of the dark matter halo in which the different galaxies are immersed. This halo dominates the dynamic
of extended galaxies with low surface brightness, and its presence is less significant in more centrally concentrated galaxies with larger surface brightness
values, e.g., \citet{MacGaughBlok1998}. 

Alternatively, the discrepancy between observable mass in galaxies and their dynamics 
can be interpreted as a direct evidence for the
failure of the current Newtonian and general relativistic theories of gravity, rather than the existence
of a dark matter component. As examples, in spiral galaxies the flat rotation 
curve has been successfully interpreted assuming a modification of Newtonian dynamics (MOND), e.g., \citet{Sanders-McGaugh2002},
or equivalently an extended Newtonian force law, e.g., \citet{mendoza11},
the projected surface density profiles and observational parameters of the local dSph galaxies are in agreement with a
description in MOND, e.g. \citet {hernandez10, McGaugh10, kroupa10}. Similarly, modified gravity approaches have been successful   
in explaining the general shape of the observed rotation curves of many dwarf and low surface brightness galaxies, e.g., 
\citet{Swaters2010}, and many other astronomical systems have been interpreted in this context, 
such as globular clusters (GCs), e.g., \citet{Sollima-Nipoti2010, Haghi2011, hernandez12b},
the relative velocity of wide binaries in the solar neighbourhood, e.g., \citet{hernandez12a}, the infall velocity of the two 
component of the  Bullet cluster, e.g., \citet{Moffat-Toth2010} and the gravitational lensing of elliptical galaxies, e.g., \citet{Mendoza2012}.

Modified gravity proposals not requiring any dark matter generically predict that for accelerations below $a_{0} \approx 1.2 \times 10^{-10} m/s^{2}$,
one should transit away from Newtonian gravity, e.g. MOND in \citet{milgrom83a}, TeVeS in \citet{bekenstein04}, QUMOND, BIMOND in \citet{zhao10}, 
the extended Newtonian force law in \citet{mendoza11} or its relativistic version in \citet{bernal11}. Regardless of the details, 
within $a<a_{0}$ scales velocities for test particles about spherical mass distributions will become independent of radius, and show
a mass scaling coinciding with that of the galactic Tully-Fisher relation of $V^{4} \propto M$.

To test the predictions of modified gravity schemes in elliptical galaxies one needs to know the kinematics and mass distribution of these systems.
Elliptical galaxies are dynamically hot systems with little or no cold gas where determining their kinematic is more difficult than for spiral galaxies. 
Their orbital structure is thought to be the result of their evolution through a complex formation process.
For determining their dynamics  several tracers have been used, such as X-ray gas, GCs and planetary nebulae (PNe).
Massive elliptical galaxies are usually surrounded by hot and low density  gas evident through its X-ray glow. A tool to determine the mass distribution
in these systems is to model their X-ray spectra to obtain density and temperature profiles for the gas, if we assume hydrostatic equilibrium for this
component, we can then obtain the mass distribution and then an estimate of the 
brightness profile to be compared to the photometrically observed one, e.g., \citet{Das2010}.

The problem of modeling the luminosity profile and kinematics of elliptical galaxies has been treated before, e.g., \citet{Kormendy2009}
modelled the brightness profiles of all known elliptical galaxies in the Virgo cluster using Sersic profiles $log I \propto r^{1/n}$, they develop 
a mechanism to calculate a realistic error in the Sercic parameters and identify departures from these profiles that are diagnostic of galaxy formation processes.
\citet{Teodorescu2011} studied the kinematics of PNe in the Virgo giant elliptical galaxy NGC 4649 (M 60), assuming Newtonian gravity, 
they conclude that the kinematics of this object are consistent with the presence of a dark matter halo around M60, this halo is almost one-half of the total
mass of the galaxy within $3R_e$. In \citet{DeBruyne2001} three-integral axisymmetric models for NGC 4649 and NGC 7097 are considered, 
concluding that the kinematic data of NGC 4649 are consistent with a dynamical model with a moderate amount of dark matter, 
\citet{Das2010} create a dynamical model of NGC 4649 using the NMAGIC code and kinematic constraints to infer  a dark matter mass fraction in 
NGC 4649 of $\sim 0.78$ at $4Re$. In \citet{Samurovic-Cirkovic2008a} and \citet{Samurovic-Cirkovic2008b} the GC dynamics of NGC 4649 are used as mass tracers and the 
Jeans equation is solved for Newtonian and MOND models, finding that both are consistent with the values of the observed velocity
profile of the galaxy, although considering the light distribution only to first approximation, as a radial power law for the surface brightness
profile.

Along the same lines, we here treat the massive elliptical galaxy NGC 4649 assuming the modified Newtonian force law formulation of \citet{mendoza11} through  
fully consistent self-gravitating models, as a test of the above ideas. The free parameters of the galactic model 
are calibrated to match the details of the observed surface brightness and projected velocity dispersion profiles. This becomes a valuable test
of the physical modelling introduced in \citet{hernandez12b} to the modelling of light and velocity dispersion profiles in Galactic
globular clusters, as the same scheme is now applied to a system which lies orders of magnitude above the globular clusters in size and mass.

We obtain equilibrium models which satisfy all observed parameters of the galaxy in question, including crucially, the flattening of the  projected velocity dispersion
evident at large radii, in the absence of any unseen dark component. Under a purely Newtonian framework, any accurate modelling of this galaxy requires
the inclusion of a substantial amount of dark matter.

In section (2) we present the modelling used to construct equilibrium models for self-gravitating  spherically symmetric stellar populations
using the modified Newtonian force law of \citet{mendoza11}. We then show in section (3) specific models satisfying all observational constraints 
available for NGC 4649. Section (4) gives our conclusions.

\section{Self-Gravitating equilibrium modeling}

In the same way as done for globular clusters in \citet{hernandez12b}
we have modeled the massive elliptical galaxy NGC 4649 as a population of self gravitating stars in a spherically symmetric 
equilibrium configuration, under a modified Newtonian gravitational force law.
For a test particle a distance $r$ from the centre of a spherically symmetric mass distribution $M(r)$, the force per unit mass is given by:

\begin{equation}
f(x)=a_{0} x \left( \frac{1-x^{n}}{1-x^{n-1}} \right).
\end{equation}

\noindent In the equation above $x=l_{M}/r$, where $l_{M}=(GM(r)/a_{0})^{1/2}$. Notice that for $a>>a_{0}$ ($x>>1$), $f(x)\rightarrow a_{0} x^{2}$ 
and we recover the standard force law, while for $a<<a_{0}$ ($x<<1$), $f(x)\rightarrow a_{0} x$ and the corresponding MONDian 
force law of $f=(M(r) G a_{0})^{1/2} /r$ ensues. The abruptness of the transition between the two limits is regulated by the index $n$,
which satisfies $n>1$.

In \citet{mendoza11} the above force law was proven to result in generalised isothermal gravitational equilibrium configurations
having well defined and finite $r_{g}$, $M$ and $\sigma$ values, radii, masses and velocity dispersion, which evolve smoothly from the 
classical virial equilibrium scaling of $M=\sigma^{2} r_{g} /G$ to the observed tilt in the fundamental plane of elliptical galaxies,
to the $\sigma^{4} =(M G a_{0})$ Tully-Fisher relation, in going from $x>>1$ to $x\sim 1$ to $x<<1$. In that paper it was also
shown that solar system constraints will not be violated, provided that the transition in eq.(1) is sufficiently abrupt, $n>4$,
and also that Newton's theorems for spherically symmetric mass distributions will continue to be valid, provided only that 
$f$ depends exclusively on the variable $x$.

Recently, in \citet{hernandez12b} we showed that constructing similarly successful models for globular clusters requires $n \geq 10$ to obtain best fits to 
brightness and velocity dispersion profiles, here we use $n=10$ to construct equilibrium models for the elliptical galaxy NGC 4649. Notice also
that given the form of the proposed force law, any other larger value of $n$ results in only marginal differences from $n=10$. Since any feasible
modified force law at the Newtonian level evidently requires the form of the force law to be fully fixed across astrophysical systems and scales,
here we do not treat $n$ as a free parameter, but use always a fixed $n=10$ value. A theoretical
grounding for the  above modified force law appeared in  \citet{bernal11}, who provided an $f(R)$ formal generalization to GR which has as its low 
velocity limit precisely the MONDian force law of eq.(1) in the $a<<a_{0}$ range.

Taking the derivative of the kinematic pressure, the equation of hydrostatic equilibrium for a polytropic equation of state $P=K\rho^{\gamma}$ is:
  
\begin{equation}
\frac{d(K\rho^{\gamma})}{dr}= -\rho\nabla \phi.
\end{equation}
In going to locally isotropic Maxwellian conditions, $\gamma=1$ and $K=\sigma(r)^2$ and we get: 

\begin{equation}
2\sigma(r)\frac{d\sigma(r)}{dr} + \frac{\sigma^{2}(r)}{\rho} \frac{d\rho}{dr}= -\nabla \phi.
\end{equation}

 since $\rho=(4\pi r^2)^{-1}\frac{dM(r)}{dr}$
the preceding equation can be written as:
\begin{eqnarray}
&&2\sigma(r)\frac{d\sigma(r)}{dr}+\sigma(r)^{2} \left[\left(\frac{dM(r)}{dr}\right)^{-1}\frac{d^{2}M(r)}{dr^{2}}-\frac{2}{r} \right]\nonumber\\
&&=-a_{0} x \left( \frac{1-x^{n}}{1-x^{n-1}} \right),
\label{eqhidro}
\end{eqnarray}

\noindent 
with $\sigma(r)$ the isotropic Maxwellian velocity dispersion for the population of stars.  
This is hence an application of the treatment developed in \citet{hernandez12b}, where we modeled eight 
GCs in our Galaxy under the same modified force law used here. In that paper we showed that using a plausible
parametric  $\sigma (r)$ function, self-gravitating models can be easily constructed to accurately reproduce all 
observational constraints.

Once a specific $\sigma(r)$ is chosen, using a numerical finite 
differences scheme and the initial condition of a fixed central density $\rho_{0}$, i.e. $M(r)\rightarrow 0$, $dM(r)/dr \rightarrow 4\pi r^{2}\rho_{0}$, 
for $r\rightarrow 0$, the  second-order differential equation (\ref{eqhidro}) can be solved.
This solution gives $\rho(r)$ and $M(r)$, volumetric profiles for the density and mass. Projecting $\rho(r)$ along one direction then
yields the corresponding $\Sigma(R)$, a projected surface density mass profile, which assuming a mass-to-light ratio, can then be compared against 
an observed surface brightness profile. We use $r$ for radial distances in 3D, and $R$ as a projected radial coordinate on the plane of the sky.

Lastly, a volume density weighted projection of $\sigma(r)$ yields $\sigma_{p}(R)$, the corresponding
projected velocity dispersion profile. Thus, it is only after solving for the complete density structure that a $\sigma_{p}(R)$
is obtained. Notice that existing observations only give  $\sigma_{p}(R)$, the density weighted projected profiles of $\sigma(r)$,
and never $\sigma(r)$ itself. To solve eq. (4) we assume a parametric form for  $\sigma(r)$, with parameters which will be
adjusted to yield full projected models in surface brightness and  $\sigma_{p}(R)$ in accordance with the properties of the observed
galaxy. We shall use:

\begin{equation}
\sigma(r)=\sigma_{1} \mathrm{exp}\left(-\frac{r}{r_{\sigma}}\right)^{m} + \sigma_{\infty}.
\label{dispvel}
\end{equation}

Notice that $\sigma_{\infty}$ will be  given directly by the observations, since at large radial distances
the effects of projection tend to zero and $\sigma(R) \rightarrow \sigma(r)$. Therefore, $\sigma_{\infty}$
can be obtained from the asymptotic value of the observed projected velocity dispersion profile for the galaxy
in question, in this case $\approx 200kms^{-1}$, following \citet{Das2011}. The remaining four free parameters,  $\rho_{0}$,
$\sigma_{1}$, $r_{\sigma}$ and $m$, will be fitted to yield projected velocity dispersion and surface brightness
profiles (following \citet{Bridges2006} we take $M/L= 8.0$) in accordance with the corresponding observed quantities.

\section{Modelling elliptical galaxy NGC 4649}

We begin this section presenting in figures (1) and (2) our best fit NGC 4649 model.
The volumetric density profile is shown in figure (1), and appears similar to a cored isothermal density distribution, with
the difference that at large radii $\rho(r)$ becomes steeper than the Newtonian $r^{-2}$. The consequence of this steepening is
a finite total mass and radius, even using asymptotically flat $\sigma(r)$ profiles e.g. \citet{milgrom1984}. This contrasts with the
Newtonian case where classical isothermal spheres of infinite extent appear.
As already pointed out by \citet{hernandez10} and \citet{mendoza11}, this finite profiles also appear for the modified gravity law we are using,
even under rigorously isothermal $\sigma(r)=\sigma_{0}$ equilibrium configurations. The vertical line shows the
radius where the condition $x=1$ is crossed, i.e., the point beyond which  $a<a_{0}$, and hence the departure of the force law used
from the Newtonian value towards the MONDian regime, consequently the point where the density profile steepens away from the $1/r^{2}$
of Newtonian gravity, towards the convergence of the MONDian regime.

\begin{figure}[!t]
\plotone{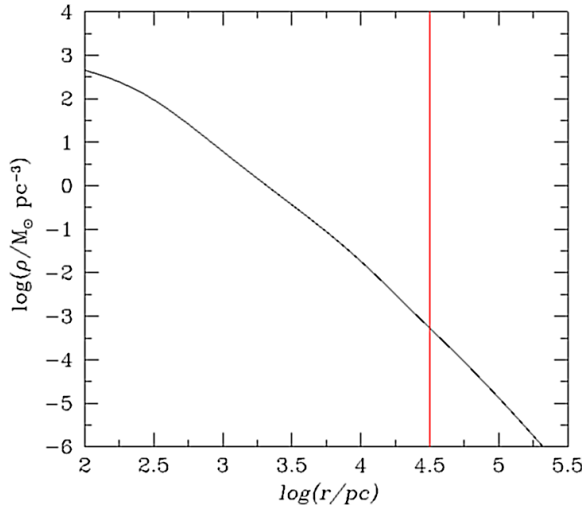}
\caption{Volumetric density profile for the best fit model for NGC 4649. The point where $x=1$, $a=a_{0}$, is shown by the vertical line.
Notice that at large radii $\rho(r)$ becomes steeper than the $r^{-2}$ of the Newtonian case, which yields a finite total
radius and mass for the model.}
\end{figure}
\begin{figure}[!t]
\plotone{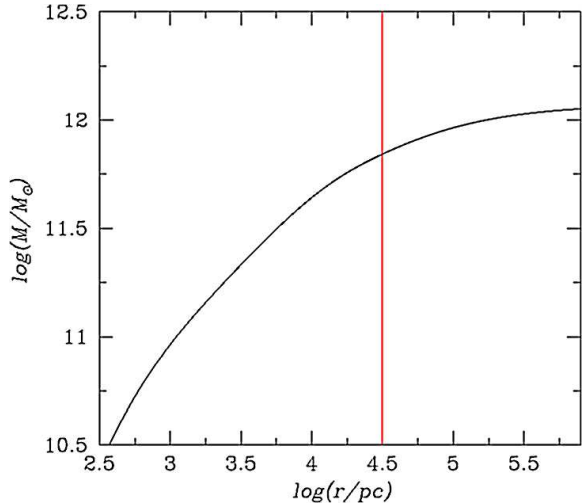}
\caption{Volumetric mass profile for the best fit model for NGC 4649.  The point where $x=1$, $a=a_{0}$, is shown by the vertical line.
Notice the convergence of the mass appearing at large radii.}
\end{figure}

The corresponding volumetric radial mass profile is given in figure (2), where the vertical line again shows the
$x=1$ threshold. This clearly shows the departure from the linear growth of the total mass within the approximately
isothermal Newtonian region, towards convergence to a finite total mass in the MONDian $a<a_{0}$ regime.  
We see that $40\%$ of the mass of the modelled galaxy falls outside of the  $x=1$ threshold. This mass appears
spread out over large areas, and hence presents much smaller surface densities than those appearing in the central
regions interior to the $a>a_{0}$, $x=1$ limit, the more easily observed Newtonian central parts.

\begin{figure}[!t]
\plotone{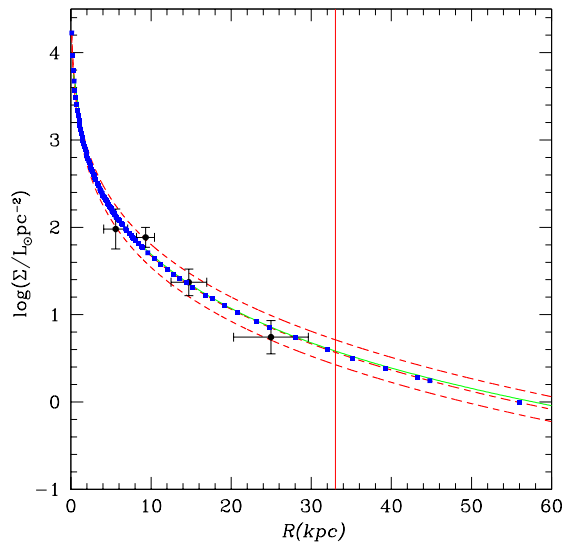}
\caption{Comparison of the resulting model projected surface brightness for NGC 4649 (continuous line) 
to the corresponding observed quantities and the best fit S'ersic function with the central (dashed line), maximum and minimum values 
consistent with the reported confidence intervals for the S'ersic parameters, flanking curves.
The squares correspond to data in \citep{Kormendy2009}, and the points with error bars for PNe data, \citep{Teodorescu2011}
Following \citet{Bridges2006} we take $M/L_V=8$, which coincides with the stellar population studies of \citet{Shen-Gebhardt2010}. 
The point where $a=a_{0}$ and the modified force law used changes from the Newtonian form to the outer MONDian behaviour, is shown by
the vertical line.}
\end{figure}

The consistency of the model is shown in figures (1) and (2); having assumed a positive isotropic Maxwellian velocity distribution function,
the physical modelling results in a mass density profile in accordance with what a Newtonian modelling would have yielded for the
inner $a>a_{0}$ region, where the force law used reproduces Newton's expression. In going to larger radii, $\rho(r)$ steepens increasingly
to eventually reach  $\rho(r)=0$ at a finite total radius, as can be seen from the convergence of the mass profile given in figure (2).
The distribution function is thus positive throughout the model, going to zero at the outer edge.

The optimum model shown was fitted to agree with the observed projected velocity dispersion and surface brightness profile for 
NGC 4649, assuming a constant  $M/L_V$ value, see below.
We take the surface brightness profile observations from photometry in \citet{Kormendy2009} and the number density 
calculated from PNe in \citet{Teodorescu2011} as scaled in \citet{Das2011}, and projected velocity dispersion measurements
from \citet{Pinkney2003} adopted from  \citet{DeBruyne2001}. Following \citet{Bridges2006} we take $M/L_V=8$, which coincides with
the central value reported by \citet{Shen-Gebhardt2010}, determined directly from stellar population synthesis models.

\begin{figure}[!t]
\plotone{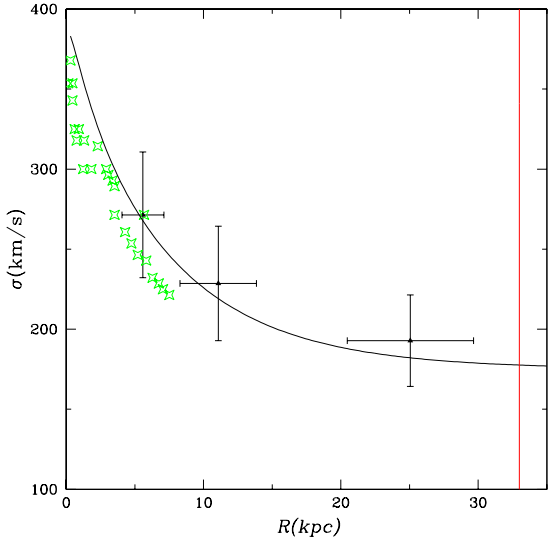}
\caption{Comparison of the resulting model projected velocity dispersion profile for NGC 4649 (continuous line)
to the corresponding observed quantities, stars for long slit observations in \citep{Pinkney2003} and the points with error bars for PNe data.
The point where $a=a_{0}$ and the modified force law used changes from the Newtonian form to the outer MONDian behaviour, is shown by
the vertical line.}
\end{figure}

\begin{figure}[!t]
\plotone{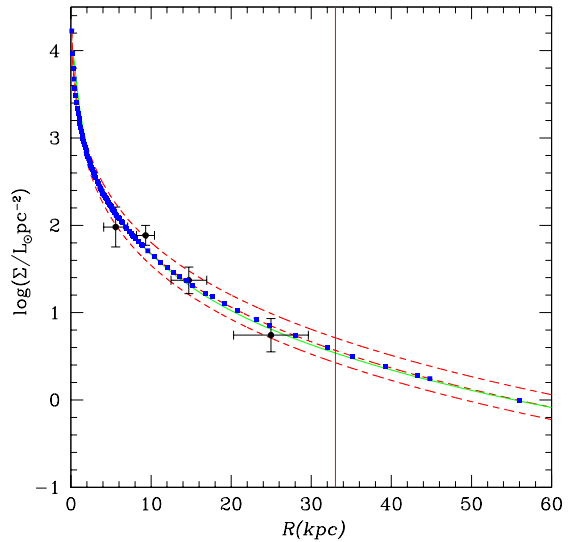}
\caption{Same as fig.(3) but using $M/L_V=6$}
\end{figure}

The final resulting surface brightness profile and its corresponding volume density weighted projected velocity dispersion profile $\sigma_{p}(R)$
for galaxy NGC 4649 are presented in figures(3) and (4) for $M/L_V=8$. 
The parameters for the best fit in eq.(5) are $\sigma_\infty=175km/s$,
$\sigma_1=230km/s$, $r_{\sigma}=8.5kpc$ and $m=1$, the central density that we have assumed is $\rho_{0}=1.5\times10^{2}M_{\odot}/pc^{3}$.

Regarding the free parameters of the model, $m, \sigma_{1}, r_{\sigma}$ and $\rho_{0}$, we began with 
$m=1$ as used in our GC study of Hernandez \& Jimenez (2012), and with a trial $r_{\sigma}$ at the effective radius of the galaxy, and a trial $\sigma_{1}$ at
the $\sigma_{p}(R=0)=\sigma_{1}+\sigma_{\infty}$ condition, we then performed a least squares fit to the well established projected surface density profile to obtain an optimal 
$\rho_{0}$ value. The projected $\sigma(R)$ profile was then visually examined, and the procedure ended if the model agreed with these more uncertain observations 
to within their reported error bands. If not, the procedure is repeated with $r_{\sigma}$ and $\sigma_{1}$ re-adjusted to yield a larger or smaller central 
$\sigma_{p}(R=0)$ value, or a faster or slower decay in the $\sigma_{p}(R)$ curve. Only at the lowest $M/L$ range were solutions with $m=1$ slightly less good than the 
ones presented, so the procedure was repeated with slight variations in $m$. We do not claim the models obtained to be unique, neither did we set out to find final 
values for our parameters which could be definitively associated to NGC 4649, merely working models intended to show the plausibility of the force law examined.
We do note however, that the full observed and Sersic fit to the projected surface density profile can be reproduced, not only within the observed uncertainties, 
but actually to an almost indistinguishable level, giving us confidence that some of the physics of the problem have been captured by the modelling presented. At the 
same time, the more uncertain projected velocity dispersion profile has also been adequately reproduced, to within the observational confidence intervals.

In the figures the vertical line indicates the point where $a=a_0$, we can see that the model accurately fits the observed projected surface 
brightness profiles of the galaxy and the central value of the measured projected velocity dispersion $\sigma_{p}(R=0)= 400km/s$, as well as the observed profiles for 
$\sigma_{p}(R)$. 

We see that both, the surface brightness and velocity dispersion profiles are in agreement with photometry and PNe observations,
to within reported uncertainties. Under Newtonian gravity this requires to invoke a large amount of dark matter to reproduce the structure and kinematics
of this galaxy, using the proposed modified force law however, it is natural that once the change in the gravitational regime appears,
$\sigma(r)$ will tend to a constant once the mass distribution has converged.

\begin{figure}[!t]
\plotone{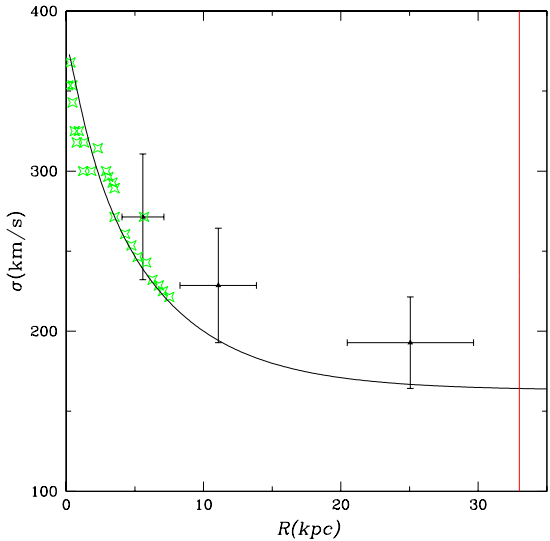}
\caption{Same as fig.(4) but using $M/L_V=6$}
\end{figure}

Although the projected surface brightness profile of the model shown in figure (3) is in excellent accordance with the observed one, a value of
$M/L_V=8$ might appear as somewhat excessive. Indeed, from figure (4) we see that the projected velocity dispersion of the model systematically
exceeds the measurements from long slit observations, and is more in accordance with the more crude PNe inferences. To test the range of values in
$M/L_V$ which the physical modelling can accommodate, we repeated the experiment, taking this time a lower value of $M/L_V=6$. The results are presented
in figures (5) and (6), which are analogous to (3) and (4). We see a similarly excellent accordance for the projected surface brightness profile
in figure (5), and a projected velocity dispersion profile in figure (6) which this time closely follows the more detailed data of the long slit 
observations, and lies slightly below the PNe inferences. The parameters of this second model are: $\sigma_\infty=163km/s$,
$\sigma_1=233km/s$, $r_{\sigma}=6.6kpc$ and $m=0.95$, with $\rho_{0}=2.3\times10^{2}M_{\odot}/pc^{3}$.

$M/L$ ratios higher than $8$ appear unphysical from the point of view of pure stellar populations, while models with values lower than $6$
result in $\sigma(R)$ curves which fall below the observed values. The range $6<M/L_V<8$ thus brackets the models which can yield
accurate simultaneous fits to both the observed surface brightness profile, and the observed velocity dispersion profile of NGC4649.

The total mass we obtain from the dynamical modeling described is consistent with what results from the integration of the surface brightness profile
assuming a constant mass to light ratio $M/L_{V}$. The mass obtained for the dynamical model to NGC 4649 is $1.09\times10^{12}M_{\odot}$, and the luminosity 
that results from integrating the $V$ band light profile is $1.36 \times 10^{11} L_{\odot}$, implying a mass of $1.08\times10^{12}M_{\odot}$ from using $M/L_{V}=8$
as determined directly from the stellar population studies of \citet{Shen-Gebhardt2010}. Also,
notice that the asymptotic value of the velocity dispersion is $195.0\pm 30.36km/s$ in consistency with the $\sigma=0.2(M/M_\odot)^{1/4}=203.9 km/s$ 
prediction of \citet{hernandez12b} for $1.08\times10^{12}M_{\odot}$, and $190.09km/s$ for the  $8.16\times10^{11}M_{\odot}$ of the $M/L_{V}=6$ case. 

Models including a degree of orbital anisotropy, constant or with radial variations, are also possible, as are ones considering also some degree
of sub-dominant rotation, as often observed in elliptical galaxies. Introducing more free functions would clearly allow even finner fits to the
observed restrictions. We have chosen not to increase the complexity of the modelling in this way, as our aim here is merely to prove that
for the simplest isotropic, non-rotating construction, reproducing all the available observational constrains is easily achieved under the
modified force tested. As already mentioned, the models presented are not intended as unique definitive fittings to any of the parameters used, merely 
as examples of plausible models showing the viability of the force law used.

\section{Conclusions}\label{ccl}

We show that for the massive elliptical galaxy  NGC 4649, fully self consistent spherically symmetric equilibrium models can be constructed using a 
modified Newtonian force law which smoothly transits from the Newtonian value when $a>a_{0}$, towards the MONDian phenomenology in the $a<a_{0}$ regime, 
which naturally satisfy all observational constraints available for the central, asymptotic and radial profiles for projected surface brightness
and velocity dispersion measurements. All this, without the need of invoking any hypothetical and as yet undetected dark matter component.

We obtain models which show a Newtonian inner region, smoothly evolving outwards towards a MONDian solution on crossing the
$a=a_{0}$ limit. The corresponding velocity dispersion profiles similarly evolve from an internal ``keplerian'' region to an
external constant velocity dispersion regime.

We show that the asymptotic value of the observed velocity dispersion profile, $\sigma_{p}(R\rightarrow \infty)$, and the total 
mass for this systems, $M$, are consistent with the generic modified gravity prediction for  
$\sigma_{p}(R\rightarrow \infty)=0.2(M/M_\odot)^{1/4}$.

\section*{acknowledgements}

The authors wish to thank an anonymous referee for comments leading to a clearer and more complete final version. 
Xavier Hernandez acknowledges financial assistance from UNAM DGAPA grant IN103011-3. Alejandra Jimenez acknowledges 
financial support from a CONACYT scholarship.


\end{document}